\documentclass[11pt,a4paper,
              ]{scrartcl}
\usepackage{graphicx}
\usepackage{subfig}
\usepackage{lcd}
\usepackage{lcd_title,ifthen}
\usepackage{mathptmx}
\usepackage{helvet}
\usepackage{amsmath}
\usepackage{color}
\usepackage{units}
\usepackage{url}
\usepackage{pbox}
\makeatletter
\def\url@myurlfontstyle{%
  \@ifundefined{selectfont}{\def\UrlFont{\sf}}{\def\UrlFont{\small\ttfamily}}}
\makeatother
\long\def\symbolfootnote[#1]#2{\begingroup%
\def\thefootnote{\fnsymbol{footnote}}\footnote[#1]{#2}\endgroup} 
\newlength{\capindent}
\newlength{\capwidth}
\newlength{\figwidth}
\newcommand{\icaption}[2][!*!,!]{\hspace*{\capindent}%
  \begin{minipage}{\capwidth}
    \ifthenelse{\equal{#1}{!*!,!}}%
      {\caption{#2}}%
      {\caption[#1]{#2}}
      \vspace*{3mm}
  \end{minipage}}
\begin{document}
\begin{titlepage}
%
%
%
\title{CLIC Muon Sweeper Design}
%
\author{A. Aloev\affiliated{1}, 
             H. Burkhardt\affiliated{1},
             L. Gatignon\affiliated{1},
             M. Modena\affiliated{1},
             B. Pilicer\affiliated{2},
             I. Tapan\affiliated{2}}

\affiliations{\affiliation[1]{CERN, Switzerland},
              \affiliation[2]{Uludag University, Bursa, Turkey}}
%
\date{\today}
%
\begin{abstract}
\noindent
There are several background sources which may affect the analysis of data and detector performance at CLIC.
One of the important background sources are halo muons, which are generated along the beam delivery system (BDS).
In order to reduce the muon background, magnetized muon sweepers have been proposed as a shielding material\,\cite{LDeacon}. 
Possible shielding magnet designs have recently been studied using the tool OPERA. 
We describe them and estimate the muon background reduction which can be achieved using the BDSIM Monte Carlo simulation code~\cite{BDSIM1,BDSIM2}.
\end{abstract}
%
\presented{Talk presented at the International Workshop on Future Linear Colliders (LCWS15), Whistler BC, Canada,
2–6 November 2015.}
%
\end{titlepage}
%
%
\section{Introduction}

The CLIC (Compact Linear Collider) project has a beam transportation line in order to deliver the beam from main LINAC to IR (interaction region) and demagnify the beam into desired parameters at the IR \cite{cdr2012}.

\begin{figure}[t]
   \centering
   \includegraphics[width=\textwidth]{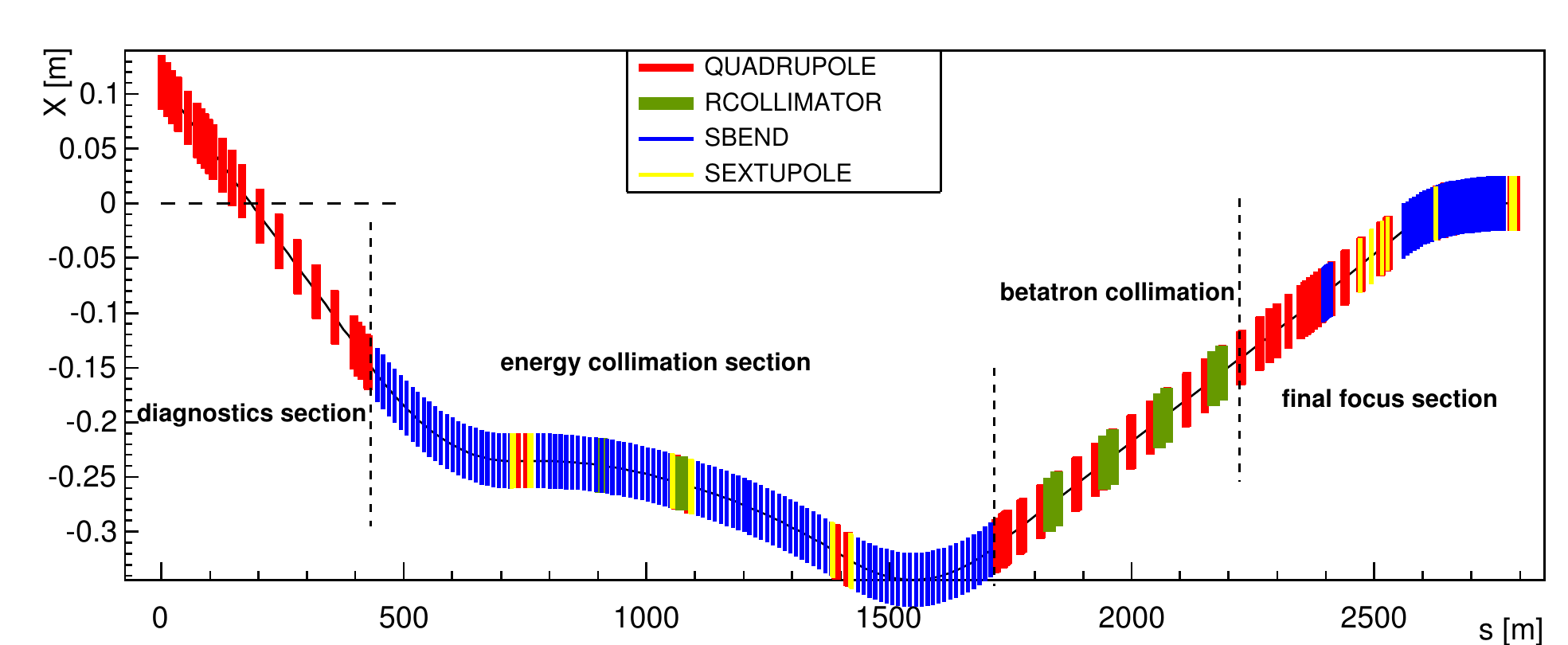}
   \caption{ Beam Delivery System layout for 3\,TeV collision. }
   \label{fig:bds}
\end{figure}

We use the layout of the BDS described in the CDR \cite{cdr2012}. The layout is shown in Figure~\ref{fig:bds}. 
After acceleration to full beam energy in the LINAC, the beam enters the diagnostics section of the BDS where the beam properties are measured and corrected.
The beam then traverses the energy collimation section, where off-momentum particles are removed. Of particular interest for this study is the third section, the betatron betatron collimation section, which is designed to remove halo particles. 
It is followed by the final focus region which has strong magnets to focus the beam to the required beam parameters at the interaction point.

Halo particles will be generated by scattering processes like beam-gas scattering. Other sources of generating and enhancing the beam halo include optics effects like mismatch and dispersion, and equipment related effects like noise or mechanical  vibrations~\cite{Fitterer2009,HTGEN}.
For the study presented here, we use halo distributions generated by the well known elastic and inelastic interactions of the beam with the residual gas.

The halo is removed in the betatron collimation section by the combination of spoilers and absorbers.
The large amplitude particles which constitute the beam halo will typically first interact with the relatively thin spoilers which are closest to the beam axis, and then be stopped by the thicker absorbers.
There is a small but finite probability that muons are produced in these interactions. Muons are much harder to stop and may reach the interaction region and deteriorate the detector performance. 
The goal of the study presented here is the mitigation of this secondary muon flux into the detector region. We propose magnetized muon shields, designed to sweep away and absorb the majority of the muons before they reach the detector region.

\section{ Muon Sweeper Design}

We propose a design with a toroidal magnetic field, which is zero on the beam axis.
We have considered two options, one based on permanent magnets and another on normal conducting magnets.

\begin{figure}[!htb]
   \centering
   \includegraphics[width=\textwidth]{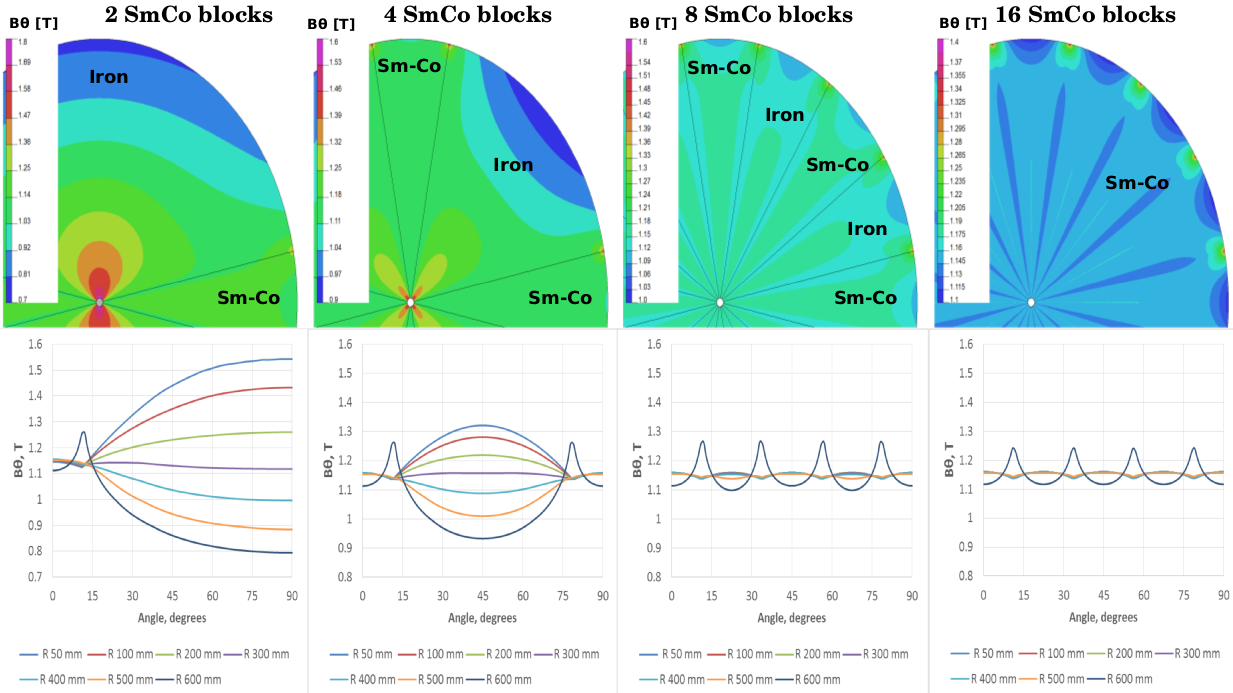}
   \caption{Magnetic field variation corresponding to angle of muon sweeper at various radius values for different number of Sm-Co blocks.}
   \label{fig:SmCo}
\end{figure}

The magnetic field homogeneity depends on number of Sm-Co blocks in the permanent magnet option.  This is shown for different numbers of Sm-Co blocks in Figure~\ref{fig:SmCo}. The average field depends on the number of Sm-Co blocks in the magnet. The magnetic field values fluctuate between 1.6\,T and 0.8\,T for 2 blocks Sm-Co option. For 16 blocks Sm-Co, the magnetic field fluctuation is reduced to values close 1.2\,T as shown in Figure~\ref{fig:SmCo}.

\begin{figure}[!htb]
   \centering
   \includegraphics[width=75mm]{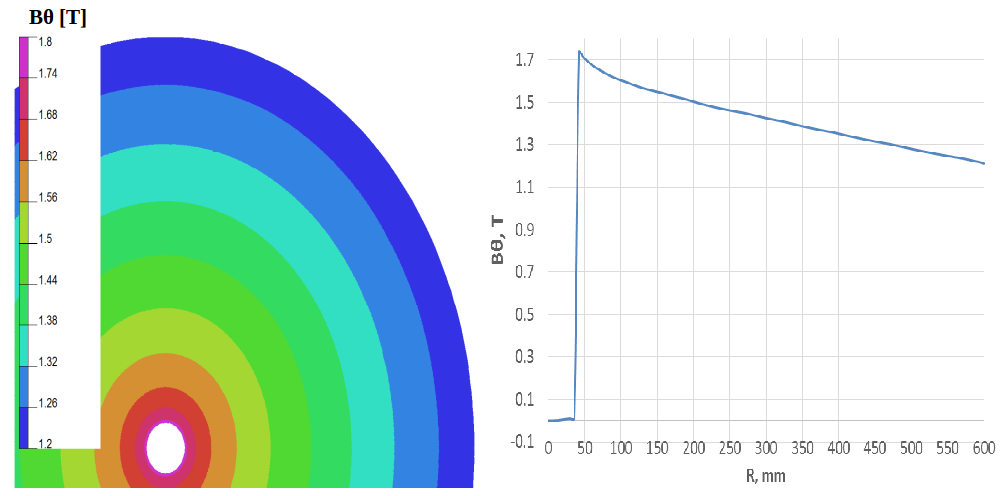}
   \caption{Magnetic field map for normal conducting coil option.}
   \label{fig:normal}
\end{figure}

The other option is based on normal conducting coils, which produce an azimuthally symmetric field shown in Figure~\ref{fig:normal}. 
The power consumption is estimated at about 120\,W for a single 2 meter section. The current density is 0.8\,A/mm$^2$ which allows for an air cooled system.
The inner and outer radius of the sweeper in all options is 4\,cm and 60\,cm, respectively.

\section{Estimation of Muon Reduction Rates}

The beam delivery system contains four sets of spoilers and absorbers in the betatron collimation section.
We perform detailed, high statistics simulations of the betatron collimation with muon production and absorption.
The muon sweepers are implemented in the simulation studies as 2.5\,m long modules with 4\,cm inner radius and 60\,cm outer radius, with 0.7\,T or 1.2\,T magnetic fields.

For the first set of simulations, four modules, each 10\,m long with 10\,cm space between them, have been placed after each vertical absorber in the betatron collimation section.
This option has 40\,m long magnetized shielding in total.

For comparison, we also re-considered the spoiler size and location from the earlier study described in Ref~\cite{LDeacon}, now based on the more realistic spoiler design presented here and using the updated version of BDSIM. In the previous study, the muon sweepers have been placed after each horizontal and vertical absorber, adding up to a total shield length of 80\,m.
We have simulated this by placing 32 of our 2.5\,m long modules after each absorber as listed in Table~\ref{table:1}.

\begin{table}[ht]
\centering
\caption{Muon sweeper length and distance to the first spoiler (YSP1) in the betatron collimation section.}
\begin{tabular}[c]{|c|c|c|c|c|}
\hline
 Muon Sweeper&Z [m] & \pbox{1.8cm}{Length [m] Ref~\cite{LDeacon}} & \pbox{3.5cm} {Modulated Length [m] (40\,m shield)} & \pbox{3.5cm}{Modulated Length [m] (80\,m shield)} \\
\hline
MS1a  & 99  & 8  & -            &3$\times$2.5\\
MS1b  & 117 & 8  & 4$\times$2.5 &3$\times$2.5\\
MS2a  & 211 & 9  & -            &3$\times$2.5\\
MS2b  & 229 & 9  & 4$\times$2.5 &3$\times$2.5\\
MS3a  & 323 & 11 & -            &4$\times$2.5\\
MS3b  & 341 & 11 & 4$\times$2.5 &4$\times$2.5\\
MS4a  & 406 & 14 & -            &6$\times$2.5\\
MS4b  & 435 & 14 & 4$\times$2.5 &6$\times$2.5\\
\hline
\end{tabular}
\label{table:1}
\end{table}

\begin{figure}[!htb]
   \centering
   \includegraphics[width=\textwidth]{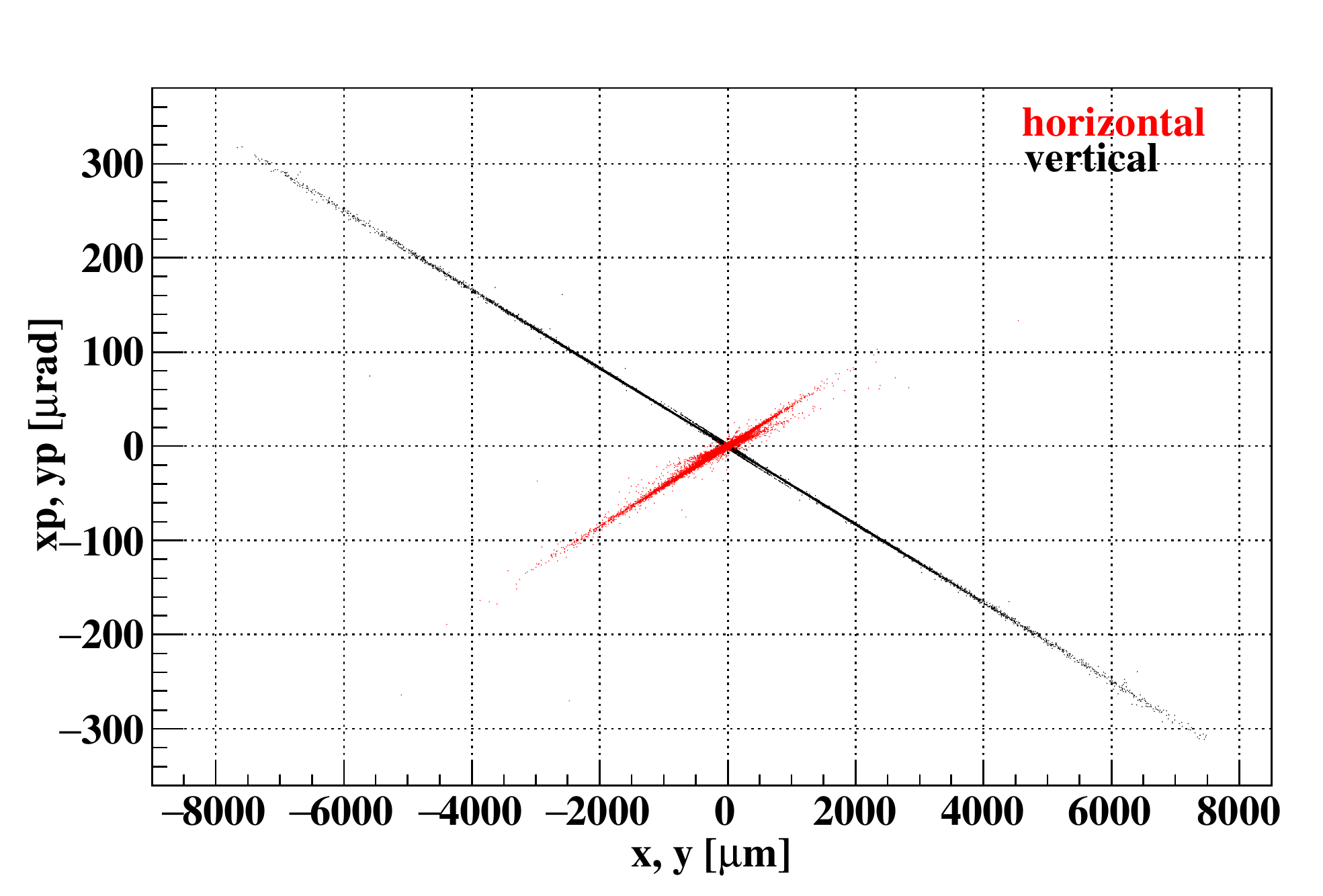}
   \caption{The estimated phase space distribution of halo particles at the first spoiler (YSP1) with HTGEN.}
   \label{fig:halodist}
\end{figure}

The muon reduction rates have been determined for each geometry for both 0.7\,T and 1.2\,T radially homogeneous magnetic fields. The halo particle distributions, both in position and energy, have been estimated by using HTGEN~\cite{HTGEN} as in Figure~\ref{fig:halodist}. The halo distribution has been used as an input parameter for primary particles hitting to the first spoiler (YSP1) in the BDSIM simulations.

\section{Conclusion}

Magnetic muon shields can significantly reduce the number of muons reaching detector region by deflecting them towards the tunnel walls.
The reduction rates depend on the magnetic field strength and the total length of the magnetized shielding.

We expect that 39.5$\pm$6.3 muons per bunch crossing would reach the detector region in the absence of magnetized shielding.
As for any absolute prediction of the muon flux to the experiments, this number depends on assumptions about the halo.
The number given here corresponds to the halo which is expected from residual gas scattering
at design vacuum parameters. It would increase if the actual rest gas pressure is higher or if the halo is enhanced by other processes.

In the study presented here, we have determined how the muon flux into the detector can be reduce by magnetized shielding.
The numbers obtained in the simulations are listed in Table~\ref{table:2}.
A factor of 4 to 6 reduction can is obtained with a 0.7\,T magnetic field for 40\,m and 80\,m long magnetized shielding, respectively. 
When the magnetic field is increased to 1.2\,T, the reduction rates increase to nearly a factor of 10. 

\begin{figure}[!htb]
   \centering
   \includegraphics[width=\textwidth]{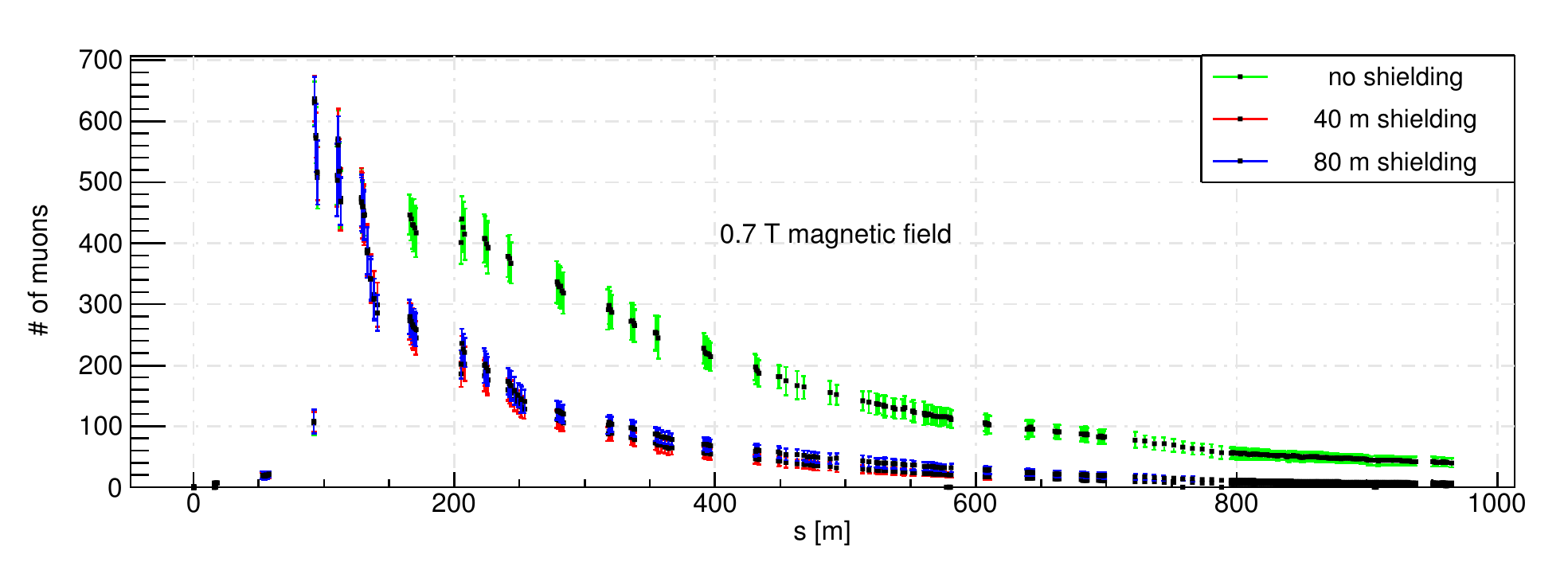}
   \caption{Changing of muon numbers from betatron collimation to interaction region without shielding and with totally 40\,m and 80\,m long shielding for 0.7\,T magnetic field.}
   \label{fig:halo07}
\end{figure}

\begin{figure}[!htb]
   \centering
   \includegraphics[width=\textwidth]{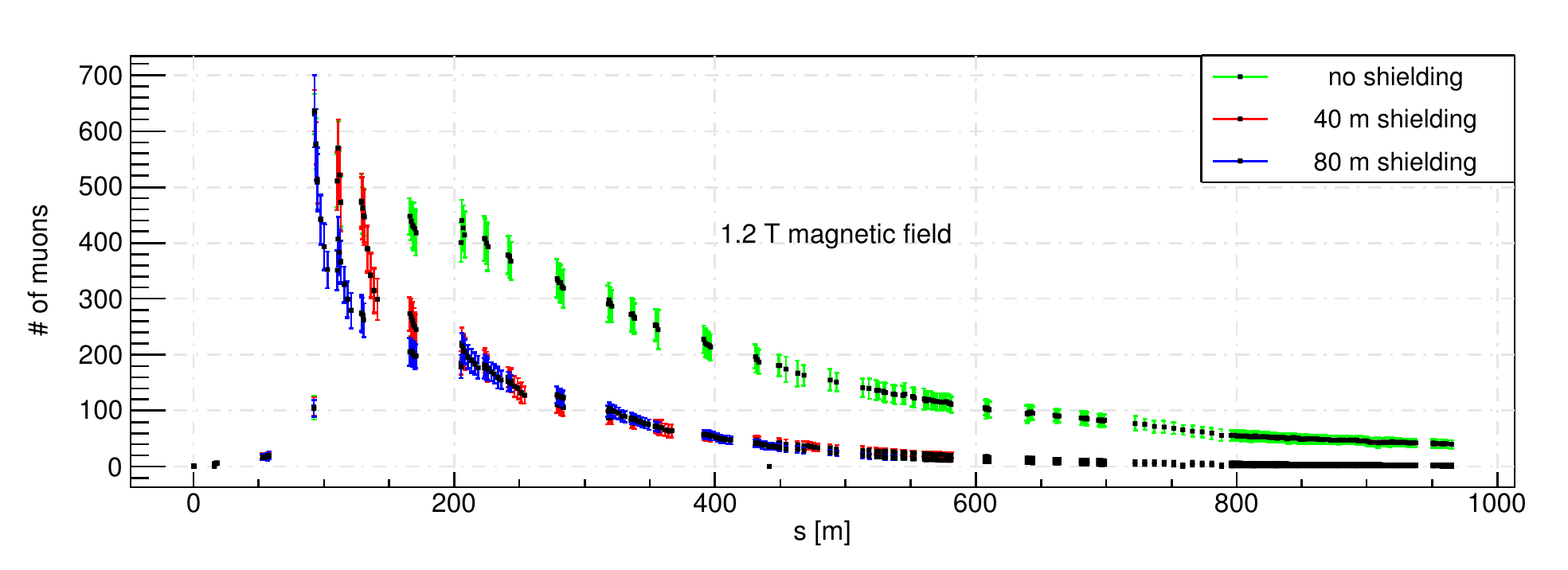}
   \caption{Changing of muon numbers from betatron collimation to interaction region without shielding and with totally 40\,m and 80\,m long shielding for 1.1\,T magnetic field.}
   \label{fig:halo12}
\end{figure}

\begin{table}[htp]
\centering
\caption{Number of muons reaching to the IR for 0.7\,T and 1.2\,T magnetic field value.}
\begin{tabular}{|c|c|c|}
\hline
        & 40\,m shielding & 80\,m shielding \\ 
\hline
0.7\,T  &  11.4$\pm$3.3  & 6.1$\pm$2.4 \\ 
1.2\,T  &   3.1$\pm$1.7  & 2.3$\pm$1.2 \\ 
\hline
\end{tabular}
\label{table:2}
\end{table}

As a result of this work, the permanent magnet and normal conducting coil options have been compared with each other by means of magnetic field homogeneity for muon sweeper. The muon sweeper parameters have been updated by taking into account the previous study with updated version of BDSIM. Roughly a factor of 10 reduction have been reached up with 1.2\,T homogeneous magnetic in the simulations. We are still having an ongoing studies to optimise muon backgrounds.

%
%
%
%

\end{document}